# Agile Process Consultation
# – An Applied Psychology Approach to Agility

*Emergent Research Forum (ERF)*

**Lucas Gren**
Chalmers and the University of Gothenburg
lucas.gren@gu.se

## Abstract

An agile change effort in an organization needs to be understood in relation to human processes. Such theory and accompanying tools already existed almost 50 years ago in applied psychology. The core ideas of Agile Process Consultation are that a client initiating a change toward more agility often does not know what is wrong and the consultant needs to diagnose the problem jointly with the client. It is also assumed that the agile consultant cannot know the organizational culture of the client's organization, which means that the client needs to be making the decisions based on the suggestions provided by the consultant. Since agile project management is spreading across the enterprise and outside of software development, we need situational approaches instead of prescribing low-level practices.

## Introduction

In organizational psychology, and more specifically in the organizational development (OD) sub-field, there is substantial knowledge in how people are affected and behave under an organizational change (Senior and Swailes 2016). It is also well known that a software development organization's transition to more agility is a change in culture that needs buy-in from all organizational levels (Iivari and Iivari 2011; Tolfo and Wazlawick 2008; Tolfo et al. 2011). The research and practice that aim at guiding agile transitions of larger companies often fail to incorporate theories and tools from organizational science and psychology (Lenberg et al. 2015a), maybe due to the fact that these technology-intensive field often mostly consists of employees with a purely technical background. However, it has lately also been shown that aspects like psychological group processes (Gren et al. 2017), interpersonal conflict (Gren 2017), and attitudes toward organizational change (Lenberg et al. 2017) are factors that need consideration in the software development context in order to optimize productivity etc., which motivate more focus on human factors in software engineering in general, and also in education (Yu 2014). There are theories and tools in organizational science and psychology that explicitly deal with such aspects in organizational change endeavors, and this paper specifically extends the theory of Process Consultation (Schein 2009; Schein 1969; Schein 1987) to agile transitions, i.e. it suggests how *Agile Process Consultation* needs to be deployed in practice. I think of the consultant as being the initiator and helper, but make no distinction if this person is coming from within the organization or function as an external consultant.

## Process Consultation

Schein (1969) idea of Process Consultation is that it underpins the assumptions of providing help, which all consultants or managers aim at offering one way or the other. One of these assumptions is that, instead of treating the managers as having good knowledge about the problem, it assumes that managers often only sense that something is wrong but have little detail of how to fix it. The consultation then consists of helping the manager to define and diagnose what is wrong and create action steps that ultimately lead to concrete changes that improve the situation. Process consultation is built on the ability to apply an active but humble enquiry, which are based on three key factors, namely: "1) Keeping clients in the driver's seat to enable them to regain status by becoming active problem solvers on their own behalf, 2) Giving them confidence that they can solve their own dilemma to some degree, and 3) Revealing as much data as possible for both the client and helper to work with" (Schein 2009). According to Schein (2009), the answer to the question of how to build and maintain the type of relationship that process consultation is, is paradoxical, because it is so simple to describe but difficult to apply many times, i.e. it is to create a communication process that will enable both the client and the consultant to figure out what is actually needed. In order to understand how Process Consultation is different, it is helpful to contrast it to other types of consultation.

## *Two Other Types of Consultation*

A common model of consultation is that a consultant provides expertise. In its simplest case, a manager defines a problem that no one within the organization or unit has expertise in, and then decides to buy that skill externally. Such problems could be in relation to technical expertise, a market analysis, or how to organize control functions of their products. The success of such an approach is dependent on four assumption: "(1) whether or not the manager has correctly diagnosed her/his own needs; (2) whether or not s/he has correctly communicated those needs to the consultant; (3) whether or not s/he has accurately assessed the capabilities of the consultant to provide the information or the service; (4) whether or not s/he has thought through the consequences of having the consultant gather such information, and/or the consequences of implementing changes that may be recommended by the consultant" (Schein 1969).

In contrast to this purchase of expertise model, Process Consultation sees the joint diagnosis of the problem as a main part of the consultation. A consultant will most of the time not have time to fully learn the client's organization since the organizational culture consists of traditions, values, and specific assumptions etc. (Schein 2010). Therefore, it is crucial that the consultant gets help in navigating through the organizational culture when help is given. In addition, it is also an assumption in Process Consultation that "any organization can improve its processes and become more effective if it can accurately locate those processes that make a difference to its overall performance /…/ the main goal of the process consultant is to help the manager to make such a diagnosis and to develop a valid action plan for her-/himself" (Schein 1969). Process Consultation is teaching diagnostic and problem-solving skills to clients, and the consultant should not attempt to actually solve the problem, that is best done from the inside.

The reasons why most change efforts fail are often attributed to the fact that the individual coworker are not given incentive enough to want to change through clear visions and empowerment (Kotter 1995), nor are their psychological reasons to resist change taken into account (Strebel 1996). For a large organizational change to happen the psychological factors need consideration. There are tools and guidelines in the psychology discourse of how such a change needs to happen (see e.g. Rollnick and Miller (1995)). Such approaches have a very strong empirical research foundation (Rubak et al. 2005) but also prescribe the process in detailed and clear steps even with suggestions of the number of workshops needed, their content, and who needs to participate. However, I will not describe these tools in further detail here, but instead focus on smaller cases of helping managers who contact a consultant and extend this to the agile case.

The second model that Process Consultation is contrasted against is the doctor-patient approach. It comprises of that one or many managers contact a consultant to "check them over." The consultant then prescribes change medicine and acts as an expert physician who knows how to treat illness. The consultant gets a great deal of power and legitimacy. However, this approach is based on the assumption that a consultant can get accurate information in order to make a correct diagnosis. The consultant would, most likely, get distorted information because coworkers' fear of retaliation. Many managers have long and extensive reports in their drawers from management consultancy firms of little practical value since the prescriptions are not understood or accepted by the "patient" (Schein 1988). The doctor-patient model depends on: "(1) whether or not the initial client has accurately identifies which person, group, or department is, in fact, "sick"; (2) whether or not the "patient" will reveal accurate information; (3) whether or not the "patient" will accept and believe the diagnosis that the "doctor" arrives at; (4) whether or not the "patient" will accept the prescription, that is, do what the "doctor" recommends" (Schein 1988).

As already mentioned, Process Consultation assumes that even if the consultant sees what is the matter in the organization, the client needs to see and define this problem her-/himself for her/him to not be defensive or disagree. The consultant only suggests remedies but the manager needs to make the decisions her-/himself. Again, Process Consultation aims at teaching diagnostic and problem-solving skills to the client in order to achieve more long-term and persistent help. Therefore, the process consultant need not be an expert of the content, but instead be an expert in helping a client do self-diagnosis in what is going on around her/him, within her/him, and between her/him and other people. The client should "remain proactive in the sense of retaining both the diagnostic and remedial initiative" (Schein 1988).

## Agile Process Consultation

The first edition of Schein's book on Process Consultation was published almost 50 years ago (Schein 1969). Already in 1969, Schein writes that: "as long as organizations are networks of people engaged in achieving some common goals, there will be various kinds of processes occurring between them. Therefore, the more we understand about how to diagnose and improve such processes, the greater will be our chances of finding solutions to the more technical problems and of ensuring that such solutions will be accepted and used by members of the organization" (Schein 1969). Today, almost 50 years later, Schein could not have been more right. Not only does the new paradigm of dealing with project in a complex and fast-changing world demands even more human interaction to function well (Cockburn and Highsmith 2001), but technological innovations within companies struggle without the understanding of human processes (like in the model-driven software development case (Selic 2003)). Therefore, I simply define Agile Process Consultation as applying Process Consultation to agile transitions, i.e. it is based on the same assumptions. Below, I will make explicit connections between these assumptions and the agile case.

What makes Agile Process Consultation different from other research on agile transformations is that, firstly, it focuses on the relationship between the consultant and the client instead of only looking at mastery of practices (e.g. (Adkins 2010). Second, it stresses the context and its culture as key aspects to fully understand, and there through help the client figure out the best solutions and initial steps forward instead of only matching the client's organization to ideal agile case (Sidky et al. 2007).

Since an agile transformation of a software organization has been shown to be a cultural change (Tolfo et al. 2011) and, therefore difficult to generalize across companies (Gren et al. 2015), I see Process Consultation as the only alternative in order to achieve such a change, which I believe is often done in practice by good agile consultants/coaches. If we look at the assumptions of Process Consultation, we can see that these are all related to the common issues agile transformations face: "(1) Clients/managers often do not know what is wrong and need special help in diagnosing what their problems actually are" (Schein 1987). Since an agile transformation spans over a number of abstraction levels in the organization, diagnosing where the bottlenecks are most often difficult. It is also naïve to think that a company would be best off throwing all their previous experience and success factors out the window because an agile approach is the Holy Grail, the end of the rainbow, and the silver bullet that removes all worries we will ever have. Actually, as accurately stated by Strebel (1994): "those who pretend that the same kind of change medicine can be applied no matter what the context, are either naïve or charlatans."

The second assumption is that "(2) clients/managers often do not know what kinds of help consultants can give them; they need to be helped to know what kinds of help to seek". Especially in software development companies where the main part of work life is in relation to technical aspects, many managers without any behavioral science background simply do not see what solutions there are to their problems. These aspects are also in relation to the third assumption: "(3) most clients/managers have a constructive intent to improve things, but they need help in identifying what to improve and how to improve it."

Since norms and organizational cultures are specific for the individual case, prescribing a software process that will work across companies and, if the implementation is pushed back, say that the company did not follow the process strictly enough, is a paradox. The core of agility is responsiveness to change (Fowler and Highsmith 2001) and following Scrum, or the popular organizational agility model SAFe (Brenner and Wunder 2015), to the letter will not deal with resistance that stem from psychological factors at the workplace, at least not if self-organization of teams is one of the goals. Understanding human processes is a must as also shown in the software development context (Lenberg et al. 2015b). Assumption four of Process Consultation is that: "(4) most organizations can be more effective than they are if they learn to diagnose and manage their own strengths and weaknesses. No organizational form is perfect; hence every form of organization will have some weaknesses for which compensatory mechanisms must be found" (Schein 1988).

The fifth assumption is in relation to the fact that a consultant cannot know the details and provide reliable course of action without the help of the client, since they know their organizational culture, which then also applies to an agile transition: "(5) a consultant probably cannot, without exhaustive and time-consuming study or actual participation in the client organization, learn enough about the culture of the organization to suggest reliable new courses of action. Therefore, unless remedies are worked out jointly with members of the organization who do know what will and will not work in their culture, such remedies are likely either to be wrong or to be resisted because they come from an outsider" (Schein 1988).

I have seen many times how agile consultants fall into the expert role and tell companies what to do, which is of very limited help when the problems are in relation to human processes. Perhaps it is a result of technically focused personnel trying to help other technically focused personnel with human processes, assumption six states that: "(6) unless the client/manager learns to see the problem for her-/himself and thinks through the remedy, s/he will not be willing or able to implement the solution and, more important, will not learn how to fix such problems should they recur. The process consultant can provide alternatives, but the decision-making about such alternatives must remain in the hands of the client" (Schein 1988). The last assumption is also critical for the agile context, since it stresses the importance of a consultant to teach the client how to diagnose and fix organizational problems. Many times, technical personnel did not receive such training beforehand in their education (Yu 2014), which means that such an investment in non-technical skills might have large positive effects, as compared to possessing no, or very little, such awareness: "(7) the essential function of Process Consultation is to pass on the skills of how to diagnose and fix organizational problems so that the client is more able to continue on her/his own to improve the organization" (Schein 1988). For an overview of human processes in organizations see Schein (1988) where he includes (1) communication processes (2) building and maintaining teams (3) problem-solving (4) group norms and culture (5) leading and influencing (6) feedback and (7) intergroup processes. The latter (point 7) is even more relevant today in relation to scaling the agile framework (see above), since intergroup processes will emerge. For further reading by the interested reader, the book does also include detailed descriptions of Process Consultation in action. Volume 2 (Schein 1987) also contains lessons for managers and consultants.

## Conclusion

Since agile project management many times fundamentally changes the culture of the organization, human processes need to be well understood (Fowler and Highsmith 2001). If agile is to be scaled to the whole enterprise the success of such a transition will heavily depend on how well managers or consultants can understand the human processes at play. There is recent evidence of this in software engineering research in relation to, for example, teams (Gren et al. 2017), and attitudes towards organizational change (Lenberg et al. 2017). Therefore, we have good reason to assume that a situational approach to agile transitions is a must and I therefore suggest the use of Process Consultation (Schein 1988) in the agile context. Further research needs to be conducted for the other parts of human processes in the software engineering domain to learn how the modern and rapidly changing agile workplace might differ from other types of organization. However, it is already clear that understanding human processes is a key success factor, also in modern agile enterprises, in relation to helping them strive for responsiveness to change in order to create value as quickly as possible. We need to situational approaches to agility since agile project management is spreading across the enterprise and outside of software development, i.e. we need a different approach to different organizations wanting to be agile instead of prescribing low-level practices. We also need research on different change paths in the agile context much like the more general work conducted by Strebel (1994).

# REFERENCES

Adkins, L. 2010. *Coaching Agile Teams: A Companion for Scrummasters, Agile Coaches, and Project Managers in Transition*. Boston, MA: Pearson Education.

Brenner, R., and Wunder, S. 2015. "Scaled Agile Framework: Presentation and Real World Example," *International Conference on Software Testing, Verification and Validation Workshops (ICSTW)*, Graz, Austria: IEEE, pp. 1-2.

Cockburn, A., and Highsmith, J. 2001. "Agile Software Development: The People Factor," *IEEE Computer* (1:11), pp. 131-133.

Fowler, M., and Highsmith, J. 2001. "The Agile Manifesto," *Software Development* (9:8), pp. 28-35.

Gren. 2017. "The Links between Agile Practices, Interpersonal Conflict, and Perceived Productivity," *21st International Conference on Evaluation and Assessment in Software Engineering*, Karlskrona, Sverige: ACM, pp. 292-297.

Gren, Torkar, R., and Feldt, R. 2015. "The Prospects of a Quantitative Measurement of Agility: A Validation Study on an Agile Maturity Model," *The Journal of Systems and Software* (107:1), pp. 38-49.

Gren, L., Torkar, R., and Feldt, R. 2017. "Group Development and Group Maturity When Building Agile Teams: A Qualitative and Quantitative Investigation at Eight Large Companies," *The Journal of Systems and Software* (124:1), pp. 104-119.

Iivari, J., and Iivari, N. 2011. "The Relationship between Organizational Culture and the Deployment of Agile Methods," *Information and Software Technology* (53:5), pp. 509-520.

Kotter, J. P. 1995. "Leading Change: Why Transformation Efforts Fail," *Harvard Business Review* (95:204), pp. 59-67.

Lenberg, P., Feldt, R., and Wallgren, L.-G. 2015a. "Behavioral Software Engineering: A Definition and Systematic Literature Review," *Journal of Systems and Software* (107:1), pp. 15-37.

Lenberg, P., Feldt, R., and Wallgren, L.-G. o. 2015b. "Human Factors Related Challenges in Software Engineering: An Industrial Perspective," in: *Proceedings of the Eighth International Workshop on Cooperative and Human Aspects of Software Engineering*. pp. 43-49.

Lenberg, P., Tengberg, L. G. W., and Feldt, R. 2017. "An Initial Analysis of Software Engineers' Attitudes Towards Organizational Change," *Empirical Software Engineering* (22:4), pp. 2179-2205.

Rollnick, S., and Miller, W. R. 1995. "What Is Motivational Interviewing?," *Behavioural and cognitive Psychotherapy* (23:4), pp. 325-334.

Rubak, S., Sandbæk, A., Lauritzen, T., and Christensen, B. 2005. "Motivational Interviewing: A Systematic Review and Meta-Analysis," *British Journal of General Practice* (55:513), pp. 305-312.

Schein, E. 1988. *Process Consultation Vol. 1: Its Role in Organisation Development*. Reading, Mass.: Addison-Wesley.

Schein, E. 2009. *Helping: How to Offer, Give, and Receive Help*. San Francisco: Berrett-Koehler Pub.

Schein, E. 2010. *Organizational Culture and Leadership*. San Francisco: Jossey-Bass.

Schein, E. H. 1969. *Process Consultation: Its Role in Organization Development*. Reading, Mass.: Addison-Wesley.

Schein, E. H. 1987. *Process Consultation. Vol. 2, Lessons for Managers and Consultants*. Reading, Mass.: Addison-Wesley.

Selic, B. 2003. "The Pragmatics of Model-Driven Development," *IEEE software* (20:5), pp. 19-25.

Senior, B., and Swailes, S. 2016. *Organizational Change*. Harlow: Pearson.

Sidky, A., Arthur, J., and Bohner, S. 2007. "A Disciplined Approach to Adopting Agile Practices: The Agile Adoption Framework," *Innovations in systems and software engineering* (3:3), pp. 203-216.

Strebel, P. 1994. "Choosing the Right Change Path," *California Management Review* (36:2), pp. 29-51.

Strebel, P. 1996. "Why Do Employees Resist Change?," *Harvard business review* (74:3), pp. 86-92.

Tolfo, C., and Wazlawick, R. 2008. "The Influence of Organizational Culture on the Adoption of Extreme Programming," *Journal of systems and software* (81:11), pp. 1955-1967.

Tolfo, C., Wazlawick, R., Ferreira, M., and Forcellini, F. 2011. "Agile Methods and Organizational Culture: Reflections About Cultural Levels," *Journal of Software Maintenance and Evolution: Research and Practice* (23:6), pp. 423-441.

Yu, L. 2014. *Overcoming Challenges in Software Engineering Education: Delivering Non-Technical Knowledge and Skills*. Hershey, Pennsylvania: IGI Global.